\documentclass{iopart}
\usepackage[utf8]{inputenc}
\usepackage{graphicx}
\usepackage{dcolumn}
\usepackage{bm}
\usepackage{tabularx}
\usepackage{booktabs}
\usepackage{amssymb}
\usepackage[colorlinks=true,linkcolor=blue,urlcolor=blue,citecolor=blue]{hyperref}
\usepackage{subfiles} 

\newcommand{\bfr}{\bm{r}}

\newcommand{\ubR}{\underline{\bm{R}}}

\newcommand{\s}{_\mathrm{{\scriptscriptstyle S}}}
\newcommand{\h}{_\mathrm{{\scriptscriptstyle H}}}
\newcommand{\xc}{_\mathrm{{\scriptscriptstyle XC}}}

\newcommand{\kB}{k_\mathrm{B}}

\begin{document}

\title[]{Training-free hyperparameter optimization of neural networks for electronic structures in matter}
\author{Lenz Fiedler$^{1,2}$, Nils Hoffmann$^2$, Parvez Mohammed$^2$, Gabriel A. Popoola$^3$, Tamar Yovell$^{1}$, Vladyslav Oles$^4$, J. Austin Ellis$^4$, Sivasankaran Rajamanickam$^3$, Attila Cangi$^{1}$}
\date{May 2022}

\address{$^1$ Center for Advanced Systems Understanding (CASUS), Helmholtz-Zentrum Dresden-Rossendorf (HZDR), D-02826 G\"orlitz, Germany}
\address{$^2$ Technische  Universit\"at  Dresden,  D-01062  Dresden,  Germany}
\address{$^3$ Sandia National Laboratories, Albuquerque, NM 87185, USA}
\address{$^4$ Oak Ridge National Laboratory, Oak Ridge, TN 37830, USA}
\ead{l.fiedler@hzdr.de and a.cangi@hzdr.de}
\vspace{10pt}
\begin{indented}
\item[]\today
\end{indented}

\begin{abstract}
A myriad of phenomena in materials science and chemistry rely on quantum-level simulations of the electronic structure in matter. While moving to larger length and time scales has been a pressing issue for decades, such large-scale electronic structure calculations are still challenging despite modern software approaches and advances in high-performance computing.
The silver lining in this regard is the use of machine learning to accelerate electronic structure calculations -- this line of research has recently gained growing attention. 
The grand challenge therein is finding a suitable machine-learning model during a process called hyperparameter optimization. This, however, causes a massive computational overhead in addition to that of data generation. 
We accelerate the construction of neural network models by roughly two orders of magnitude by circumventing excessive training during the hyperparameter optimization phase. We demonstrate our workflow for Kohn-Sham density functional theory, the most popular computational method in materials science and chemistry.
\end{abstract}

\section{Introduction}
%
The electronic structure of matter can be viewed as nature's glue \cite{kurth_role_2000} that binds atoms together into condensed systems like molecules and solids, thereby shaping the diversity of chemical systems and materials that surround us. A wide variety of materials characteristics including structural, elastic, and response properties are determined by the electronic structure \cite{martin_electronic_2004}.
Pressing questions from industry and society such as finding better materials for photovoltaics, identifying more efficient catalysts, designing future battery technologies, and discovering materials with novel properties are linked directly to the electronic structure of matter.

Electronic structure calculations are indispensable for complementing experimental investigations in materials science, and the need for ever more accurate and particularly efficient calculations is unbroken.
Currently, the most widely used electronic structure method is Kohn-Sham density functional theory (DFT) \cite{hohenberg_inhomogeneous_1964,kohn_self-consistent_1965, mermin_thermal_1965} due to its balance of accuracy and computational efficiency. Under the assumption of the Born-Oppenheimer approximation \cite{born_zur_1927}, by employing the Kohn-Sham formalism, and the ever growing variety of exchange-correlation functionals \cite{toulouse_review_2021}, DFT enables electronic structure calculations for a large range of systems. 

However, large-scale simulations at system sizes reaching millions of atoms, typically encountered in state-of-the-art molecular dynamics simulations, remain a final frontier for DFT. While DFT methods may possess favorable scaling properties compared to other ab-initio approaches, DFT calculations can usually only be performed for hundreds and up to a few thousand atoms. This applies especially to dynamical settings \cite{dziedzic_practical_2020} or to systems exposed to high temperatures \cite{karasiev_liquidliquid_2021}. Larger calculations can only be accomplished in special cases with enormous computational cost and time \cite{nakata_large_2020}, thereby rendering large-scale investigations infeasible. 
The traditional pathway to circumventing these technical restrictions relies on algorithmic advances in software which leads to computationally more efficient calculations. Alternatively, approximate models such as average atom models \cite{dharma-wardana_liquid-liquid_2020, massacrier_reconciling_2021, callow_first-principles_2021} are applied for otherwise unattainable calculations. While they have a smaller computational overhead, they sacrifice accuracy.

However, a drastically different option is to combine the power of machine learning (ML) with DFT data. This emergent field of research is growing fast \cite{fiedler_deep_2021,wei_machine_2019,gubernatis_machine_2018,carleo_machine_2019,schmidt_recent_2019}. It currently focuses on extracting application-specific information from DFT data sets \cite{ulissi_address_2017,schutt_how_2014,ramakrishnan_big_2015} and constructing interatomic potentials \cite{deringer_machine_2017,sosso_neural_2012,morawietz_density-functional_2013} for molecular dynamics simulations based on Gaussian process regression \cite{bartok_gaussian_2010}, ridge regression \cite{schmidt_predicting_2017}, or neural networks (NNs) \cite{smith_ani-1_2017}.

Here, we however focus on using ML to directly tackle the electronic structure problem in terms of the Kohn-Sham equations \cite{kohn_self-consistent_1965}.
Pioneering efforts include kernel ridge-regression \cite{snyder_finding_2012, brockherde_bypassing_2017,tsubaki_quantum_2020} 
and deep NN \cite{tsubaki_quantum_2020} models for predicting the electronic density. Based on these efforts, NN models for predicting the local density of states (LDOS) have recently been developed \cite{chandrasekaran_solving_2019,ellis_accelerating_2021}. These models are more general than those based solely on the electronic density. They replace traditional DFT calculations by enabling direct access to both the electronic structure and related observables, such as the total energy.

\begin{figure*}[ht]
    \centering
    \includegraphics[width=0.95\textwidth]{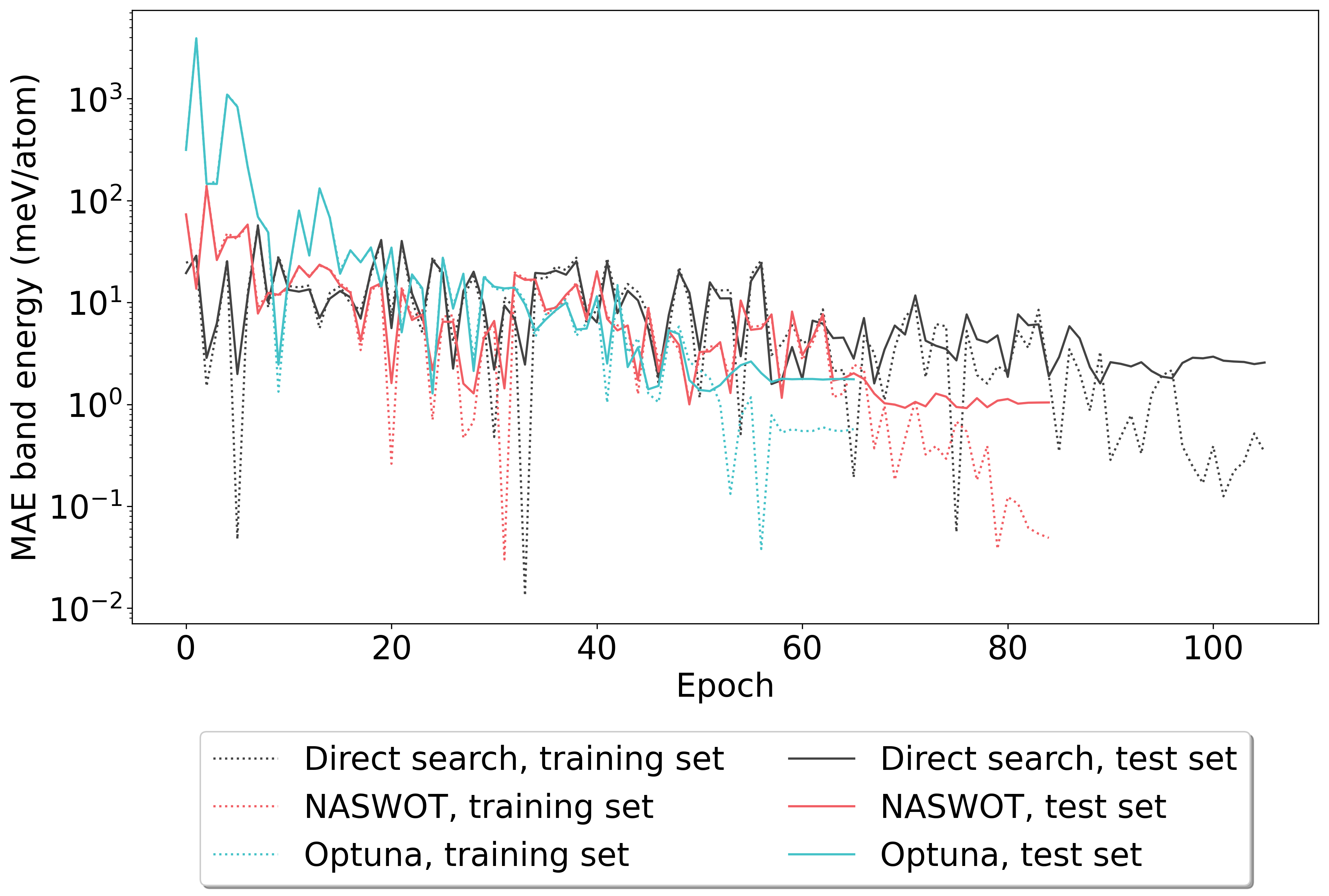}
    \caption{Training curves for three different model architectures and optimization parameters determined by three different hyperparameter optimization techniques. The loss in term of the physically motivated band energy on both the training (dashed) and test set (solid line) per epoch is shown. The grey curves correspond to direct search, the baseline for our manuscript as given in Ref.~\cite{ellis_accelerating_2021}, the blue curves to the Optuna \cite{akiba_optuna_2019} library and the red curves to the NASWOT method \cite{mellor_neural_2020}. Drastic differences can be observed, with NASWOT and Optuna outperforming the direct search approach in overall accuracy. Further, the Optuna model shows the best generalizability, as quantified by the smallest difference in test vs.~training loss.}
    \label{fig:intro_figure}
\end{figure*}

Despite these pioneering efforts \cite{brockherde_bypassing_2017}, a general-purpose workflow for automated ML applications which can tackle electronic structures has yet to come. The principle challenge is to overcome the massive computational overhead due to hyperparameter optimization, which is a general problem for ML methodologies \cite{hutter_beyond_2015}. Determining suitable hyperparameters for a NN surrogate model governs model accuracy and generalizability, as is illustrated in Fig.~\ref{fig:intro_figure} for different hyperparameter optimization techniques discussed in this manuscript. It constitutes a challenging task, even if the underlying electronic structure of the system is well understood (e.g.~metals at room temperature, which are investigated in this work).

\begin{figure*}[ht]
    \centering
    \includegraphics[width=0.95\textwidth]{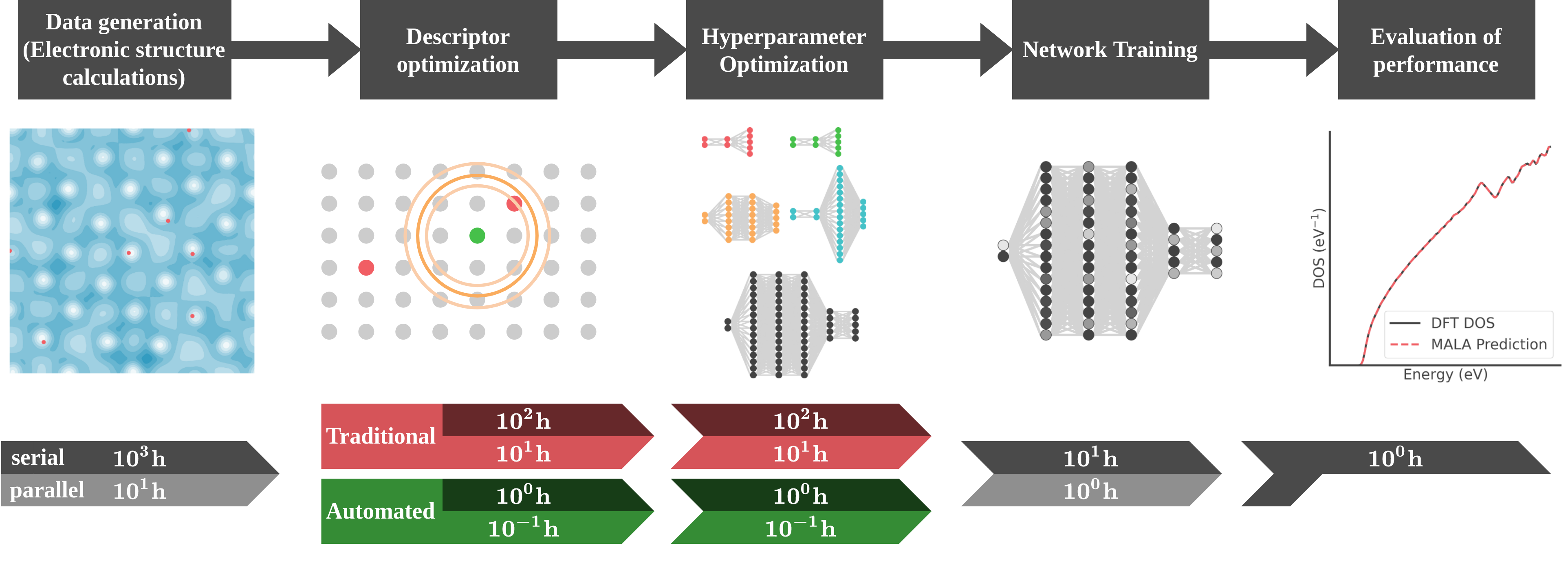}
    \caption{Schematic overview over the proposed workflow. Constructing surrogate models in the traditional fashion (red) is contrasted with our accelerated workflow (green). Here, $h$ denotes CPU/GPU hours for serial execution or wall-time for parallel execution. Timings are estimated with justifications given in the Supplementary Note 3. The pictograms show from left to right: a contour plot of the electronic density of an aluminum cell for $z=0$, with atoms close to the plane projected onto it in red; the descriptor calculation around a grid point (green dot) incorporating information of adjacent grid-points (grey) and atoms (red) within a certain radius (orange circles), which has to be optimized (light orange dots); different candidate networks; tuning of network weights; comparison of actual and predicted outputs (here, density of states).}
    \label{fig:full_workflow}
\end{figure*}

A large amount of compute time must be dedicated to relatively long and inaccessible training and optimization processes, in which high-fidelity data sets have to be constructed, suitable hyperparameters have to be identified, and models have to be constructed.
The major bottleneck in this process is not due to the computational overhead of data generation, as efficient DFT codes and automation frameworks  \cite{huber_aiida_2020,larsen_atomic_2017, uhrin_workflows_2021} exist. While DFT calculations may require large amounts of computational power, they can be completed in a reasonable amount of \textit{wall-time} by means of efficient parallelization and independent individual calculations. Instead, the principal time constraint for the creation of surrogate ML models lies in the hyperparameter optimization, which requires the repeated training of potentially suitable NNs, as illustrated in Fig.~\ref{fig:full_workflow}. The computational cost of NN training quickly becomes excessive, especially when considering a wide range of hyperparameters. While the training costs may be amortized by subsequent accelerated dynamical simulations, they render ML surrogate models prohibitive for many applications. 

In this paper, we tackle this problem by providing a highly efficient and automated hyperparameter optimization workflow for generating ML surrogate models of electronic structures. It speeds up this process by two orders of magnitude (see Fig.~\ref{fig:full_workflow}) and comprises two central components -- a \emph{training-free score} and a \emph{descriptor surrogate metric}. 
We adapt the recently developed technique on neural architecture search without training (NASWOT) \cite{mellor_neural_2020} as a \emph{training-free score} for electronic structures which does not require any NN training up until an optimal set of hyperparameters has been identified. It correlates well with the accuracy of a NN, as we demonstrate in a comprehensive comparison with state-of-the-art hyperparameter optimization techniques (see Fig.~\ref{fig:hyperparameter_comparison_timings}). 
Furthermore, we also introduce the average cosine similarity distance (ACSD) as a highly efficient \emph{descriptor surrogate metric} for finding optimal descriptors for particle-mesh data (see Fig.~\ref{fig:snap_analysis_cutoff_radius} and Fig.~\ref{fig:snap_anaylysis_components}).

We hence provide the basis for automated ML \cite{chauhan_automated_2020} workflows for modeling electronic structures. The resulting software framework, MALA \cite{cangi_mala_2021}, enables researchers to construct DFT surrogate models without extensive knowledge in ML or access to leadership-class computational infrastructure. We thus pave the way towards accessible and large-scale electronic structure calculations driven by ML.

\section{Methods}

\subsection{Density Functional Theory}
Density functional theory (DFT), which we exclusively use in the Born-Oppenheimer approximation \cite{born_zur_1927}, is based upon the Hohenberg-Kohn theorems \cite{hohenberg_inhomogeneous_1964} and the Kohn-Sham formalism \cite{kohn_self-consistent_1965}. In the context of this work, we work within finite-temperature DFT \cite{mermin_thermal_1965}, i.e., $\tau>0\,K$. Within DFT, a system consisting of $N_e$ electrons and $N_i$ ions is treated. The ionic positions $\ubR$. For ease of notation, we drop the parametric dependence on $\ubR$ and energetic contributions due to ion-ion interactions in the following. Also note that here and below we adopt Hartree atomic units, i.e., $\hbar = e = m_e = a_0 = 1$. 

The central quantity in DFT is the electronic density $n$, which is defined via the Kohn-Sham wave functions $\phi_j$ as 
\begin{equation}
    n(\bfr) = \sum_j f^\tau(\epsilon^\tau_j)\, |\phi_j(\bfr)|^2 \; ,
\end{equation}
with $f^\tau$ being the Fermi-Dirac distribution function and $j$ running over all Kohn-Sham wave functions included in a particular DFT calculation. Their number has to be adjusted according to $\tau$, as large temperatures lead to thermal excitations. The Kohn-Sham wave functions constitute a non-interacting auxiliary system restricted to reproduce the interacting electronic density and are governed by one-particle Schrödinger-like equations, often referred to as Kohn-Sham equations
\begin{equation}
    \left[-\frac{1}{2}\nabla^2 + v^\tau\s(\bfr)\right]\phi_j(\bfr) = \epsilon^\tau_j \phi_j(\bfr) \; .
\end{equation}
Here, $v^\tau\s(\bfr)$ refers to the Kohn-Sham potential, which is determined self-consistently and incorporates the interaction with ions as well as a mean-field description of electron-electron interaction, while $\epsilon^\tau_j$ are the energy eigenvalues of $\phi_j(\bfr)$.
DFT calculations are performed in an iterative fashion, seeking to identify an electronic density which minimizes the total (free) energy
\begin{equation}
    A^{\mathrm{BO}}_\mathrm{total}[n] = T\s[\phi_j] - \kB \tau S\s[\phi_j] + E\h[n] + E^\tau\xc[n] + E^{ei}[n] \label{eq:KohnShamEnergy} \; ,
\end{equation}
with the kinetic energy $T\s$ and entropy $S\s$ of the non-interacting system, the electrostatic interaction terms of the electronic density with itself $E\h$ and with the ions $E^{ei}$. All energetic contributions not included in these terms are absorbed into the exchange-correlation functional $E^\tau\xc$, thus keeping the framework formally exact. Practical calculations are enabled by appropriate approximations for $E^\tau\xc$, such as the LDA-PW91 \cite{ceperley_ground_1980} or the PBE functional \cite{perdew_generalized_1996} for $\tau=0\,K$; the development of functionals for $\tau>0\,K$ is an area of active research \cite{groth_ab_2017,karasiev_accurate_2014,brown_exchange-correlation_2013}. \\
In the DFT surrogate models we construct, the central role of the density is replaced by the local density of states $d(\epsilon, \bfr)$ (LDOS), defined via 
\begin{equation}
    d(\epsilon, \bfr) = \sum_j {|\phi_j(\bfr)|}^2 \delta(\epsilon-\epsilon^\tau_j) \label{eq:ldos} \; . \\ 
\end{equation}
For practical calculations, $\delta(\epsilon-\epsilon^\tau_j)$ has to be approximated numerically, and the way this approximation is performed is one of the hyperparameters of the DFT surrogate workflow. The advantage of using the LDOS is that it gives direct access to the total free energy of a system, as both the electronic density, as well as the electronic density of states can be given in terms of the LDOS as 
\begin{eqnarray}
    D(\epsilon) &= \sum_j \delta(\epsilon-\epsilon^\tau_j) = \int d\bfr \; d(\epsilon, \bfr) \; ,  \\
    n(\bfr) &= \sum_j f^\tau_j {|\phi_j(\bfr)|}^2 = \int d \epsilon\;  f^\tau(\epsilon) d(\epsilon, \bfr) \; , 
\end{eqnarray}
and Eq.~(\ref{eq:KohnShamEnergy}) can be expressed in terms of $n$ and $D$. More precisely, $D$ can be used to calculate the electronic entropy and band energy $E_b$ and with these the total free energy is determined 
\begin{eqnarray}
	A^\mathrm{BO}_\mathrm{total}[d] = &E_b[D[d]] - \kB\tau S\s[D[d]] - E\h[n[d]] \nonumber  \\ &+ E^\tau\xc[n[d]] - \int d\bfr\, v^\tau\xc(\bfr)n[d](\bfr)  \; . \label{eq:EnergyFromLDOS}
\end{eqnarray}

\subsection{Neural network based surrogate models}

Drawing on Eq.~(\ref{eq:EnergyFromLDOS}, one can construct surrogate models for DFT simulations by finding a suitable way to approximate the LDOS, which at each point in (simulated) space $\bm{r}$ is a vector in the $\epsilon$ dimension. For this task, we employ a NN. 

NNs, which we will refer to as $M$, are powerful regression models consisting of layers of so called neurons or perceptrons \cite{rosenblatt_perceptron_1957}. Each neuron performs a linear operation using weights $\mathbf{W}$ and biases $\bm{b}$ on provided inputs $\bm{x}$ and thereafter applies a non-linear activation function $\sigma$, yielding intermediate outputs $\bm{y}$ as
 \begin{equation}
     \bm{y} = \sigma(\mathbf{W}\bm{x} + \bm{b})\,,
 \end{equation} 
which serve as input for subsequent neurons. In connecting multiple layers containing an arbitrary number of neurons, in principle any function can be approximated, as long as the NN is tuned on an appropriate amount of data in a process called training \cite{hornik_multilayer_1989,hornik_approximation_1991}. Such a training process is usually performed using gradient methods in an iterative fashion, based on back-propagation \cite{goodfellow_deep_2016}. For the training of a NN, the available data is divided into a training data set (to calculate the gradients), a validation data set (to monitor NN accuracy during training), and a test data set (unseen during training and used to verify performance after training is completed). Each pass of the entire training data set through the network is labelled an {\it epoch}. NN performance is highly dependent on the correct choice of hyperparameters $\boldsymbol{\lambda}$ that characterize both architecture and training policy for a NN, such as the number and width of individual layers. We use feed-forward NNs, in which each neuron of a layer is connected to all neurons of subsequent layers. NNs were constructed using the PyTorch library \cite{paszke_pytorch_2019} and trained on a single GPU.

We use NNs to build a \textit{per-grid-point} model, and map a vector at each point in space to a corresponding LDOS vector. To this end we employ Spectral Neighborhood Analysis Potential (SNAP) \cite{thompson_spectral_2015,wood_extending_2018,wood_data-driven_2019,cusentino_explicit_2020} descriptors $B(\bfr, j)$ that are calculated as functionals of $\ubR$ and encode local information around a grid-point. At each point $\bm{r}$, a feed-forward NN $M[\boldsymbol{\lambda}]$ then performs a mapping of the type
\begin{equation}
    \tilde{d}(\epsilon, \bfr) = M(B(j, \bfr))[\boldsymbol{\lambda}] \; , \label{eq:NNPassing}
\end{equation}
that depends on a set of hyperparameters $\boldsymbol{\lambda}$. Thereafter the LDOS is post-processed to compute the desired output quantities, like energies, forces, etc.~analytically.

Within this work, all NNs have been trained with an algorithm consistent with Ref.~\cite{ellis_accelerating_2021}. For all networks, \textit{early-stopping} was employed, i.e.~training was halted once improvement of validation accuracy starts to stagnate. How long such stagnation has to be present for the algorithm to stop is one of the hyperparameters we tune with the algorithm outlined below. More information on technical details of network training can be found in Supplementary Note 1.

\subsection{Hyperparameter optimization}

Hyperparameter optimization aims to find a set of hyperparameters $\boldsymbol{\lambda}$, such as the number or width of layers in an NN that minimizes a loss term $L$. Generally, this is an optimization problem where we need to minimize the loss
\begin{equation}
    L =L\left(Y, \tilde{Y}(X)[\boldsymbol{\lambda}]\right) \; , \label{eq:PrincipalMetric}
\end{equation}
where $Y$ is the function we seek to model and $\tilde{Y}$ is its prediction based on input data $X$. Solving this complex optimization problem is a formidable task and forms a computational bottleneck in all ML applications \cite{hutter_beyond_2015}.
For applications in the realm of electronic structures, we choose
\begin{equation}
    L =L\left(d(\epsilon, \bfr), \tilde{d}(\epsilon, \bfr), N\right) \; , \label{eq:PrincipalMetric2}
\end{equation}
where $d$ denotes the LDOS, $N$ the total number of atomic snapshots, and $\tilde{d}$ is calculated via Eq.~(\ref{eq:NNPassing}). 

Multiple loss metrics are conceivable in Eq.~(\ref{eq:PrincipalMetric2}). Since we are not interested in the LDOS itself, $L$ is chosen to be the mean absolute error (MAE) of some quantity calculated via the LDOS. This quantity is chosen to be the total free energy $A$, and the loss metric becomes
\begin{equation}
    L = \frac{1}{N}\sum_{i=1}^{N} \Big{|} A\big[d_i\big]-A\big[M(B_i)[\boldsymbol{\lambda}]\big]\Big{|} \; , \label{eq:FinalMetric}
\end{equation}
where $d_i$ is the reference LDOS obtained from DFT for an atomic configuration labelled by the index $i$, while the NN aims to reproduce $d_i$ using the SNAP descriptors $B_i$.
During hyperparameter optimization, $A$ is usually replaced by the band energy $E_b$ due to better computational accessibility, and Eq.~(\ref{eq:FinalMetric}) is evaluated only at the very end.

Hyperparameter optimization is usually performed by choosing a \textit{candidate} model sampled from the entirety of potential hyperparameters and then optimizing this model, which constitutes a \textit{trial}. Various hyperparameter optimization techniques differ in the way they propose candidate models and refine potential hyperparameter guesses based on the information provided. Within this work, we investigate several state-of-the-art methods, namely
\begin{enumerate}
    \item \textbf{Direct search} approach \cite{hooke__1961, lewis_direct_2000}, as performed in Ref.~\cite{ellis_accelerating_2021},
    \item Tree-structured Parzen Estimator (TPE) \cite{bergstra_algorithms_2011} as implemented in the software library \textbf{Optuna} \cite{akiba_optuna_2019},
    \item \textbf{Optuna} coupled to a \textbf{NASWOT-based pruner},
    \item Orthogonal Array Tuning \textbf{OAT} \cite{zhang_deep_2019} and 
    \item \textbf{NASWOT} \cite{mellor_neural_2020}, both with \textbf{fixed} and \textbf{optimized} training schemes.
\end{enumerate}

The direct search approach serves as a baseline that we have used in prior work. Therein, one optimizes one hyperparameter at a time while holding the others fixed and training multiple candidate networks per hyperparameter, thus progressively generating more suitable models. In a similar fashion, the TPE algorithm implemented in Optuna improves model performance progressively by constructing a model to determine the relationship between the hyperparameter values and the performance observed in trials. To increase robustness, we perform each trial multiple times and report an averaged result to Optuna, so as not to trap the algorithm in local minima due to unfavorable network initializations. Optuna can be coupled to a \textit{pruner}, i.e.~a metric that discards unpromising trials. As we demonstrate below, NASWOT can be employed in such a fashion. 

NASWOT and OAT  are in principle exhaustive and sample the entire search space presented. In the case of OAT this is done via constructing orthogonal arrays, a concept often employed in experimental design. The orthogonal arrays yield a compact list of potential hyperparameter combinations, that in principle covers the search space at drastically reduced computational cost. After measuring the performance of all trials in such a list, a range analysis is performed to extract the optimal set of hyperparameters from it. One important limitation of this approach is that orthogonal arrays do not exist for any arbitrary number of (hyper)parameters \cite{beder_note_2017}, thus one has to either restrict or artificially inflate the search space. 

NASWOT achieves a full sampling of the search space by measuring the performance of individual candidate NNs without training the NN at all, thus allowing for a large number of trials to be performed. One can simply test all possible combinations of hyperparameters. The NASWOT method relies on the correlation between the input and output of a NN. Here, this correlation is quantified in terms of a Jacobian $\mathbf{J}$ which is defined as the derivative of the predicted LDOS w.r.t. the SNAP descriptors. It assigns a score to a given NN upon initialization defined as 
\begin{equation}
    S_\mathrm{NASWOT} = \sum_i^{N_\mathrm{batch}}\left[\log(\sigma_{J,i}+k)+{(\sigma_{J,i}+k)}^{-1}\right] \; . \label{eq:NASWOT_foundation}
\end{equation}
Here, $N_\mathrm{batch}$ denotes the size of a subset of the training data passed through the NN via Eq.~(\ref{eq:NNPassing}), $\sigma_{J,i}$ the eigenvalues of the correlation matrix obtained from the Jacobian, and $k$ a small parameter which ensures numerical stability. We then calculate the NASWOT mean score across five network initializations in terms of Eq.~(\ref{eq:NASWOT_foundation}) as
\begin{equation}
        S'_\mathrm{NASWOT}=\overline{S}^T_\mathrm{NASWOT}+\sigma\left(S^T_\mathrm{NASWOT}\right) \; , \label{eq:NASWOT_in_practice}
\end{equation}
where $\overline{S}^T_\mathrm{NASWOT}$ denotes the mean and $\sigma\left(S^T_\mathrm{NASWOT}\right)$ the standard deviation across the $T$ individual scores. This is done to increase robustness of $S^T_\mathrm{NASWOT}$ w.r.t.~network initialization. $S'_\mathrm{NASWOT}$ then serves as a metric to determine how well an untrained NN can distinguish between given data points. The underlying assumption is that a network performing well at this task upon initialization will also yield a high accuracy after training. Assuming $S'_\mathrm{NASWOT}$ is sufficiently correlated with the prediction accuracy of a NN after training, Eq.~(\ref{eq:NASWOT_in_practice}) provides a computationally inexpensive means for performing hyperparameter optimization. It replaces the usual loss metric, such as in Eq.~(\ref{eq:FinalMetric}), which is computationally heavy, because it needs to be computed after training. It is to note that while the loss in our framework needs to be minimized, the NASWOT score needs to be maximized. 

However, NASWOT can only optimize the architecture of a NN. All hyperparameters related to model optimization cannot be treated by NASWOT, as no model is optimized. We therefore test NASWOT in two settings - one we deemed \textbf{fixed} (i.e.~optimization parameters are set to fixed values) and one called \textbf{optimized}, where after having determined the optimal network architecture via NASWOT, optimization related hyperparameters are identified via a reduced Optuna study. Tab.~\ref{tab:hyperparameter_methods} gives an overview over the hyperparameter optimization techniques discussed here. 

\begin{table}
    \centering  
    \caption{Overview over different hyperparameter techniques employed throughout this work, the hyperparameters they are capable of optimizing, stopping criteria, and parallelization capabilities.}\label{tab:hyperparameter_methods}		
    \begin{tabularx}{\textwidth}{XXX}
    	\toprule
    	\textbf{Method} &  \textbf{Optimized Hyperparameters} & \textbf{Parallelization} \ \\\midrule
        Direct search  & Architecture and Optimization. & Upper limit for parallelization is number of potential values per hyperparameter. \\
        Optuna (Tree-Structured Parzen Estimator (TPE) & Architecture and Optimization. & Parallelization in principle unlimited, in practice too many parallel instances lead to more uninformed trials and slow convergence.   \\
        Optuna coupled to a NASWOT-based pruner & As above. & As above. \\
        Orthogonal array tuning (OAT)  & Architecture and Optimization. & Upper limit for the parallelization is the number of rows in the orthogonal array. \\
        NASWOT (fixed) & Architecture. & Upper limit for the parallelization is the number of trials. \\
        NASWOT (optimized) & Architecture and Optimization. & Architecture: as NASWOT (fixed); Optimization: as Optuna. \ \\\bottomrule
    \end{tabularx}
\end{table}


\subsection{Data and code availability}
All machine-learning experiments and post-processing analysis have been carried out with the MALA code, version 1.1.0. Training data and benchmark models for aluminum experiments can be found in Ref.~\cite{ellis_ldossnap_2021} and for beryllium experiments in Ref.~\cite{fiedler_ldossnap_2022}. Code used to conduct hyperparameter optimization as well as the relevant models can be found in Ref.~\cite{fiedler_scripts_2022}.

\section{Results}

Based on Eq.~(\ref{eq:PrincipalMetric2}), we have carried out large-scale hyperparameter optimizations in order to identify the most suitable techniques for automated DFT surrogate model generation. Our results below are divided into two categories -- (1) those for optimizing hyperparameters determining the NN architecture and (2) methods for choosing the most suitable descriptors. The former technique can be applied to any ML workflow which deals with a mapping of vector quantities, while the latter highlights how physical insight can be used to accelerate modeling specifically in the materials science domain.

\begin{figure*}[ht!]
    \centering
    \includegraphics[width=0.8\textwidth]{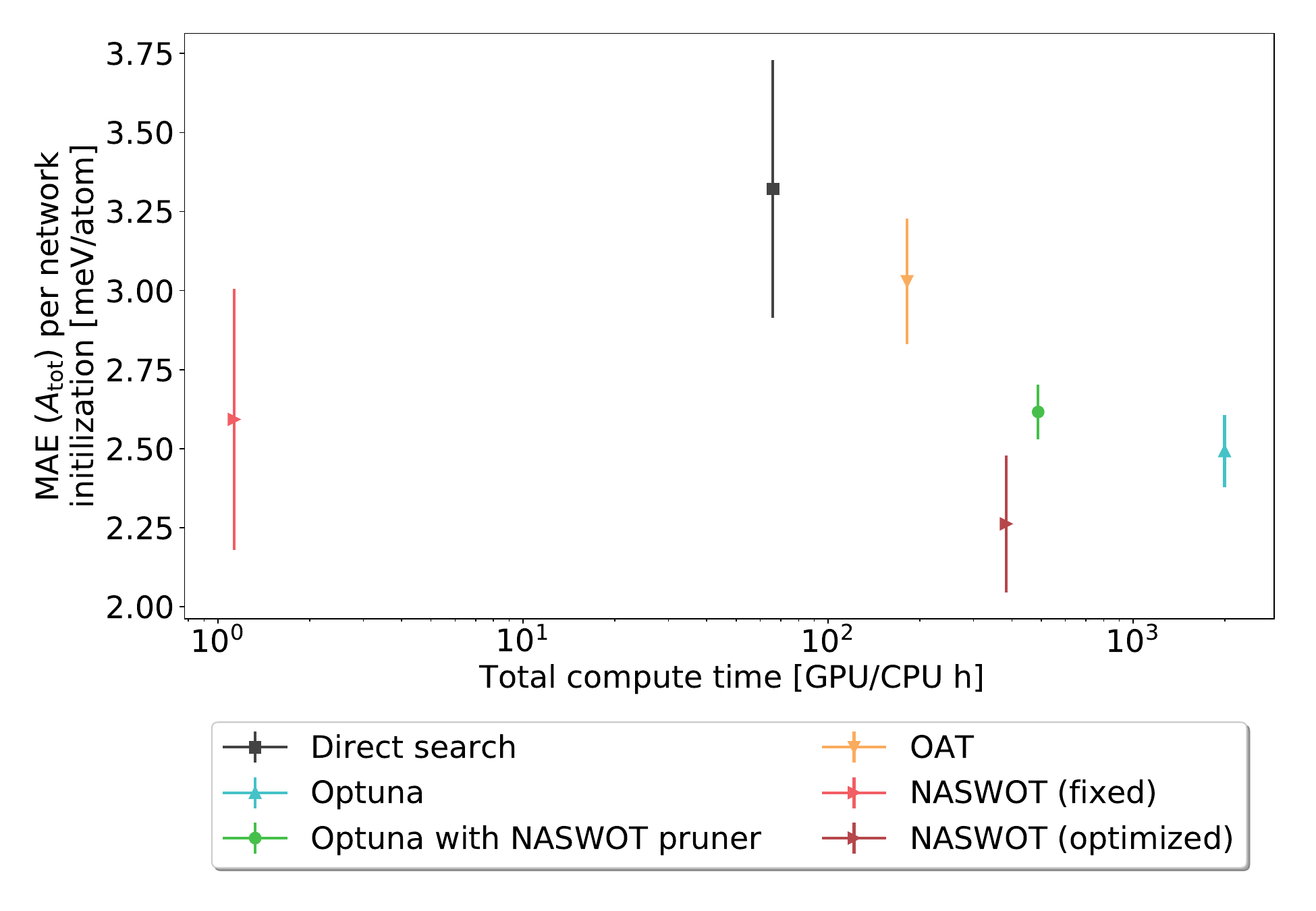}
    \caption{The \emph{training-free score}, which we implemented as the NASWOT mean score (red), provides a speed up by two orders of magnitude compared to state-of-the-art hyperparameter optimization methods. We report the mean absolute errors (MAE) across 10 atomic configurations vs. the computational demand. Markers indicate the average MAE and antennas the standard deviation over these MAE. The reported optimal NN architectures are obtained by completing five training cycles within each hyperparameter optimization method. The same methodology is employed in the subsequent figures.}
    \label{fig:hyperparameter_comparison_timings}
\end{figure*}

\subsection{Training-free hyperparameter optimization score}
We compare the NASWOT mean score with state-of-the-art hyperparameter optimization methods and highlight its utility as a superior alternative to conventional hyperparameter optimization schemes that are based training. These optimizations share the common goal of identifying a NN architecture and training routine with a minimal prediction error in the shortest amount of time. Ideally, the accuracy of a NN should be independent of the NN initialization, while the inference time should be minimal.

In our assessment of hyperparameter optimization techniques, we consider a simulation cell containing 256 aluminum atoms at room temperature (298 K) and ambient mass density (2.699 g/cc) \cite{ellis_ldossnap_2021}. This system represents the complexity of learning electronic structure data while still being computationally tractable for extended studies~\cite{ellis_accelerating_2021}.

To better quantify the accuracy of each hyperparameter optimization method, we train the model identified as optimal five times, each time using a different network initialization. Then, for each initialization, inference was performed across 10 atomic configurations that had previously been used in Ref.~\cite{ellis_accelerating_2021}. Using these prediction results, the MAE was calculated according to Eq.~(\ref{eq:FinalMetric}). In doing so, we get five MAEs, one for each network initialization per hyperparameter optimization method. We use this information to assess both accuracy and robustness of the listed hyperparameter optimization techniques. Details on the hyperparameter ranges used in these experiments are provided in Table 1 of Supplementary Note 1.

Our central result is illustrated in Fig.~\ref{fig:hyperparameter_comparison_timings}. It shows the MAE vs. total compute time for a NN identified by the considered hyperparameter optimization techniques, while also quantifying the robustness of the NN by showing the spread for the MAE across different network initializations as antennas for the data points. It demonstrates that our \emph{training-free score} implemented as the NASWOT mean score (red) provides a speed-up of two orders of magnitude while maintaining high accuracy comparable to the other methods. The average accuracy of the NASWOT NN is better then obtained from the direct search (black) and OAT (orange). Only the Optuna-based methods outperform it slightly, but at the price of a massive computational cost.
Generally, it is quite evident that for the most part, increasing computational time yields more accurate NNs, and, as quantified by the standard deviation over the inference accuracy, more robust training routines. However, the pure Optuna approach (blue) identifies an NN with excellent inference and training performance at a cost almost two orders of magnitude higher then a direct search. 
The NASWOT mean score yields a NN with relatively large variances with respect to NN initializations. These are explained by the fact that NASWOT itself has no means to adjust parameters such as the learning rate. Therefore a fixed choice is required. Performing a small Optuna study afterwards (brown) drastically reduces this variance, while at the same time introducing an additional computational overhead. The length of such an Optuna study varies depending on demands for accuracy and availability of computational resources. Chiefly, a huge reduction in compute is enabled by our \emph{training-free score} if a larger variance between network initializations can be tolerated. Even more accurate NNs become attainable with subsequent and slightly larger Optuna studies. 
 
\begin{figure}[ht]
    \centering
    \includegraphics[width=0.95\columnwidth]{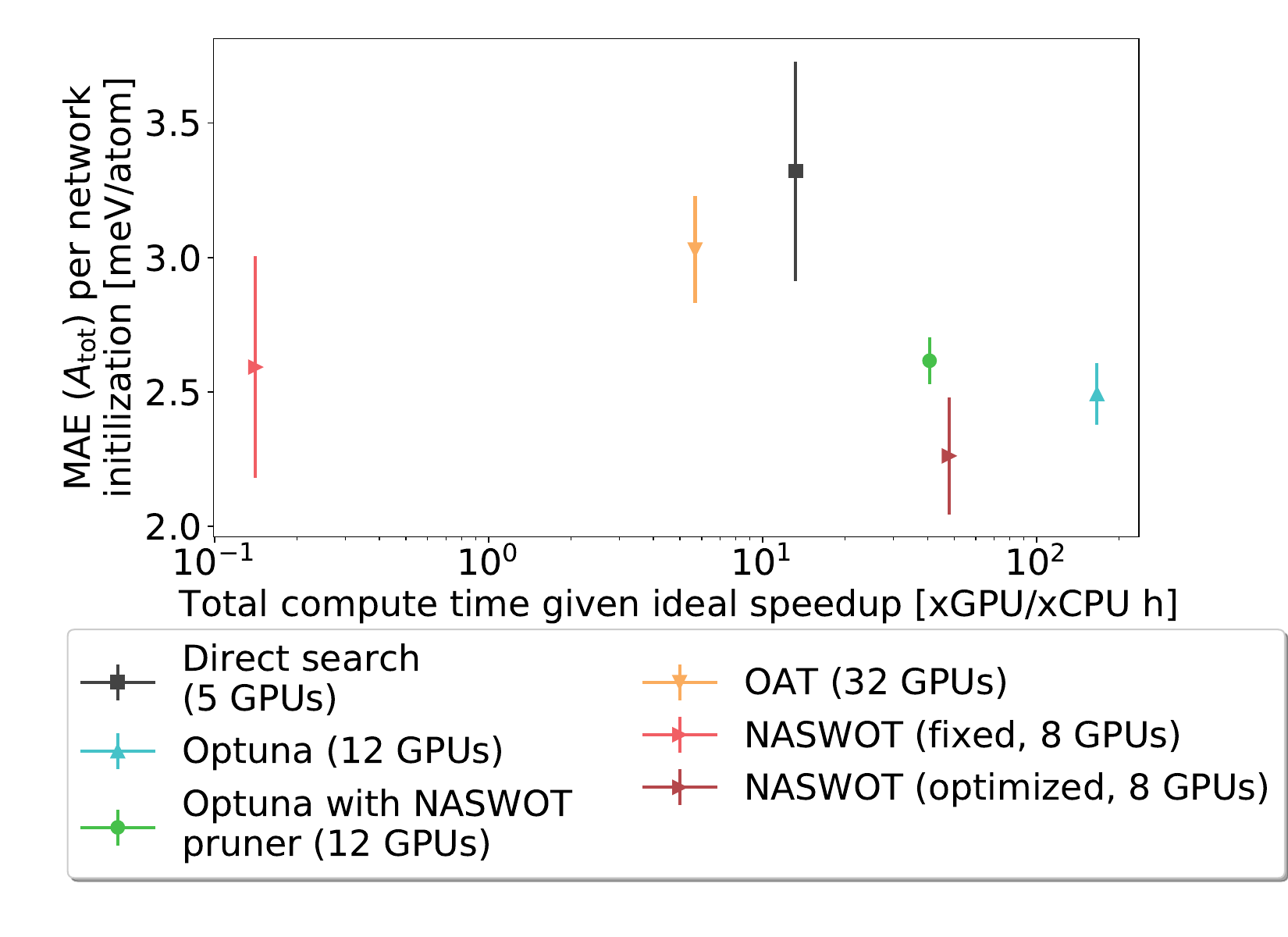}
    \caption{The \emph{training-free score} (red) also provides a speed-up by two orders of magnitude when the time to solution is considered. Reported MAE (across 10 atomic configurations) vs. computational demand for different hyperparameter optimization techniques discussed above, with maximum expected speed-ups; actual speed-ups may differ slightly. Note that there are upper limits for parallelization. If too many GPUs are used for the Optuna parallelization, trials will be too independent, leading to slower convergence. The upper limit for OAT is the number of trials, since these are completely independent of each other. However, only a serial version of OAT was implemented for now. For the direct search, one GPU can be assigned per choice. For NASWOT there is no principal upper limit, but the same number of eight GPUs as in the Optuna study was chosen.}
    \label{fig:hyperparameter_comparison_timings_parallel}
\end{figure}

Naturally, one is often not directly interested in the total core hours, which measure both time and resources, but rather in the total time-to-result. To this end, Fig.~\ref{fig:hyperparameter_comparison_timings_parallel} assesses speed-ups due to parallelization of the hyperparameter studies. 
While this improves the performance across all hyperparameter optimization techniques, no drastic changes in their order can be observed. OAT has a more favorable scaling behavior then the direct search algorithm leading to a smaller runtime. Using Optuna to optimize training hyperparameters of the NASWOT NN yields the most accurate network in runtime of two days. Yet the principal result remains the same as in Fig.~\ref{fig:hyperparameter_comparison_timings}: NASWOT outperforms the other hyperparameter techniques again by two orders of magnitude. 
The central result displayed both in Fig.~\ref{fig:hyperparameter_comparison_timings} and Fig.~\ref{fig:hyperparameter_comparison_timings_parallel} is the first step towards automated ML surrogate model generation for electronic structures.

\begin{figure}[ht]
    \centering
    \includegraphics[width=0.95\columnwidth]{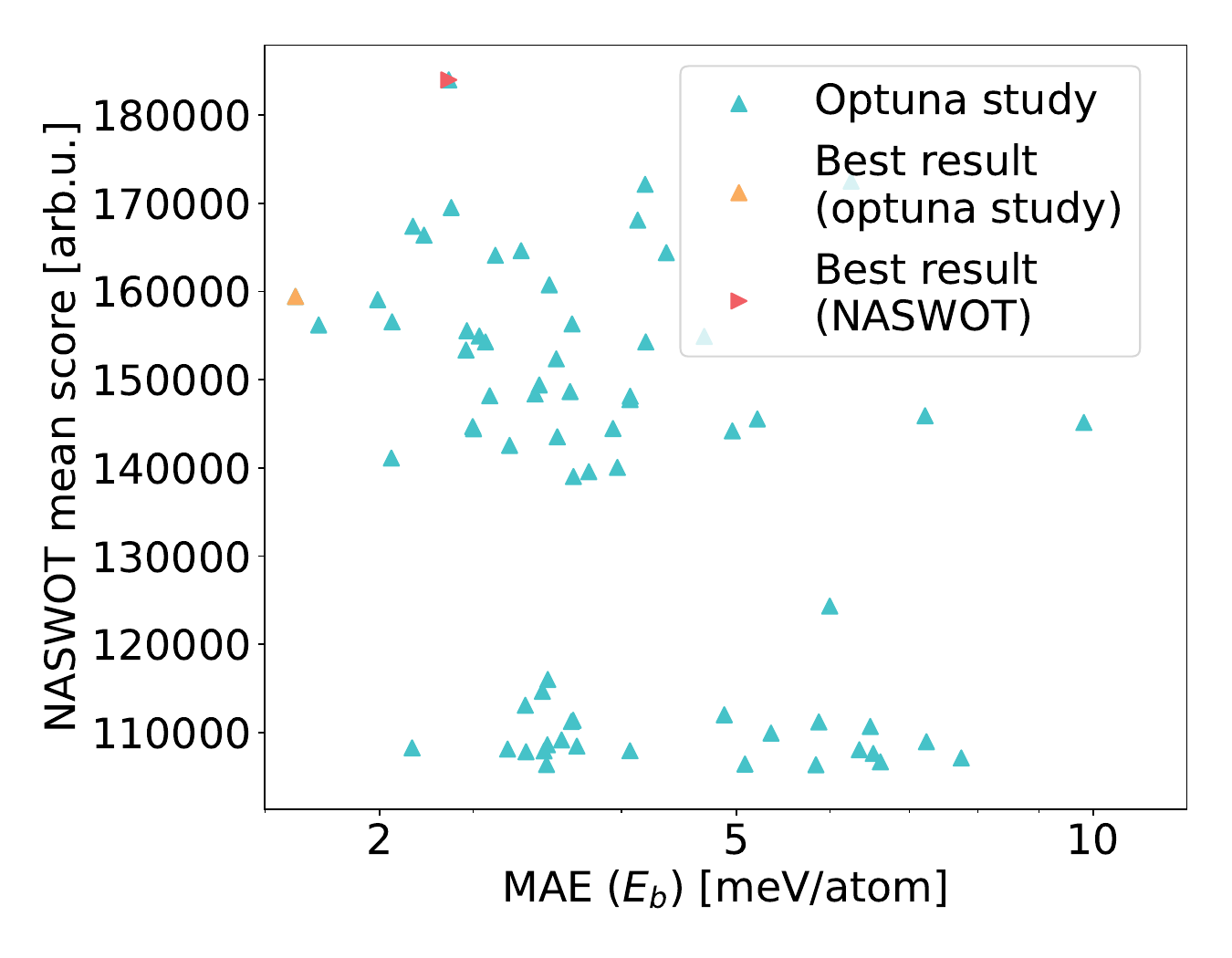}
    \caption{Training-free score (NASWOT mean score) vs.~band energy prediction on the validation set of the Optuna study.  The optimial NN architecture identified by NASWOT and Optuna are marked in red and orange, respectively.}
    \label{fig:naswot_on_optuna}
\end{figure}

The NASWOT method relies on the correlation between the training-free score calculated upon initialization of an NN and the performance of said NN after training. Thus, the performance of NASWOT as shown in Fig.~\ref{fig:hyperparameter_comparison_timings} and Fig.~\ref{fig:hyperparameter_comparison_timings_parallel} itself is not sufficient to assess whether it is a reliable tool for automated surrogate model generation. Care has to be taken to ensure that such results are actually representative of a useful underlying correlation and not caused by, e.g., an imprecise problem statement. To this end, the correlation between the NASWOT mean score and the NN performance is analyzed in Fig.~\ref{fig:naswot_on_optuna}.
The data basis for this figure is the Optuna study itself. Therefore training related hyperparameters can vary between different trial NNs. In doing so, it is ensured that the performance score assigned through this analysis is actually representative of suitable NN candidates. In order to save computation time, only the band energy rather than the total free energy was calculated for each candidate NN. The resulting comparison still provides the necessary insight, as it was shown in Ref.~\cite{ellis_accelerating_2021} that errors in the total free energy are dominated by errors in the band energy. It can be seen both visually and from the calculated Kendall $\tau$ coefficient \cite{kendall_new_1938} of $-0.318$ that the quantities shown in Fig.~\ref{fig:naswot_on_optuna} are negatively correlated, meaning that large NASWOT scores relate to small NN errors, as is the expectation. While NASWOT and Optuna do not agree in their choice for the optimal NN architecture, the overall difference is not drastic; the optimal Optuna NN is still within the 20 best NNs according to the NASWOT score, while the NASWOT result yields a reasonable accuracy in the band energy. Based on these results, the NASWOT mean score is a suitable metric for identifying \textit{a} NN capable of learning the LDOS to the desired accuracy.

Furthermore, the two clusters of points in Fig.~\ref{fig:naswot_on_optuna} suggest that NASWOT can be used to optimize hyperparameter studies which are based on Optuna. A pruner can be constructed, that discards all candidate NNs below a certain threshold score. When treating new materials, such a threshold will be unknown beforehand and has to be estimated during runtime from preceding trials. Yet for first tests this simple implementation suffices. As shown in Fig.~\ref{fig:hyperparameter_comparison_timings_parallel}, 
the resulting hyperparameter optimization technique labelled as Optuna with NASWOT pruner (green) provides high accuracy and little uncertainty while coming at a significantly lower computational cost than the direct use of Optuna. However, one has to keep in mind that for unknown systems, additional computation time has to be included to accommodate for the incremental construction of the pruning threshold. Thus, such an approach is only a viable alternative to the NASWOT algorithm if computation time is not scarce, but yet not as abundant as necessary for a full Optuna study. Naturally, other pruning algorithms could be used to perform a similar task. OAT has been found to give an intermediate performance compared to the aforementioned methods. OAT is in principal highly parallel, giving it the best performance after NASWOT (provided enough GPUs are available), while at the same time outperforming a traditional direct search in terms of accuracy.

\begin{figure}[ht]
    \centering
    \includegraphics[width=0.9\columnwidth]{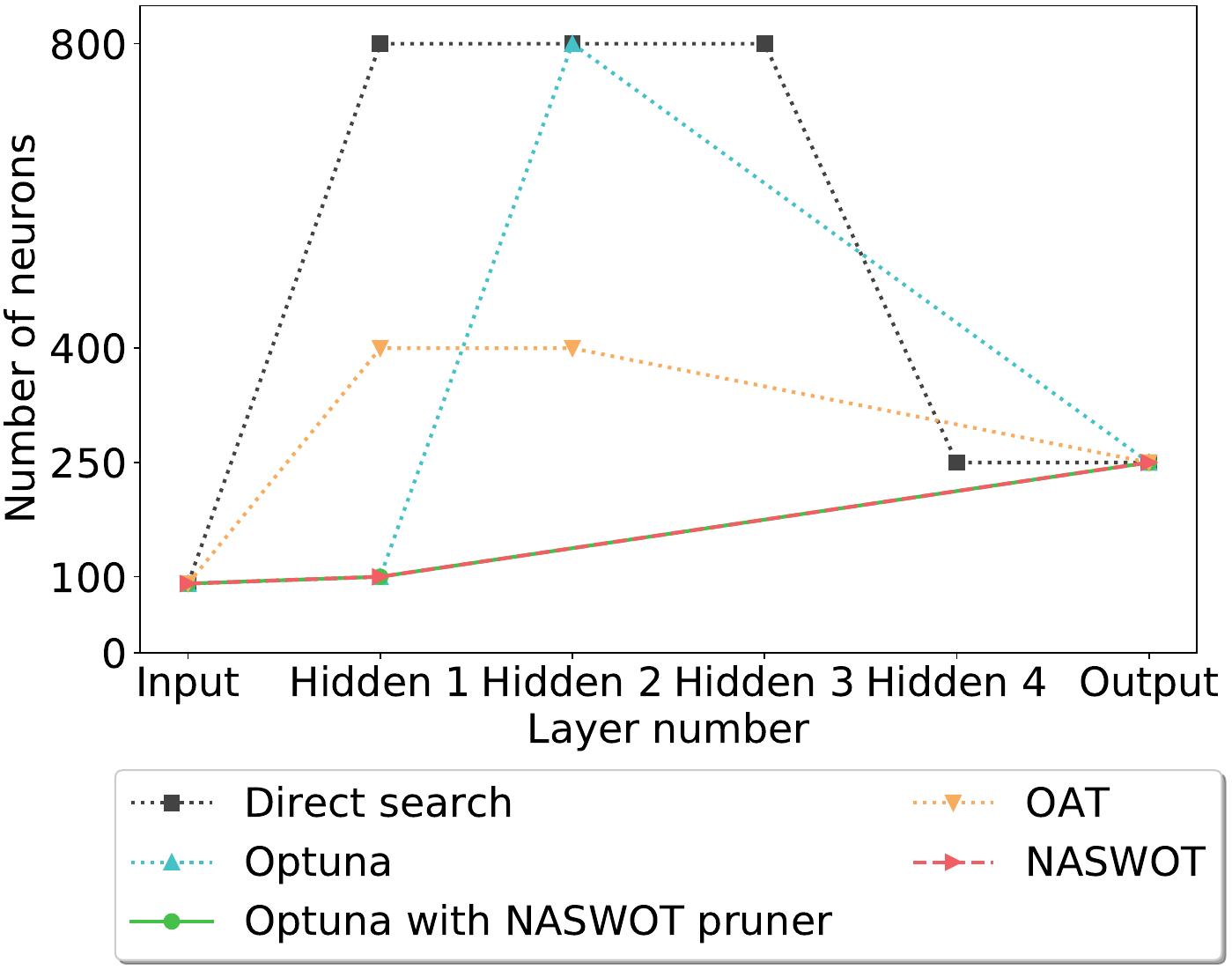}
    \caption{NN architectures identified as optimal by different hyperparameter optimizations.}
    \label{fig:network_sizes}
\end{figure}

A final, important aspect of hyperparameter optimization is the identification of an \textit{efficient} NN architecture. For initial studies, the size of a NN might not be crucial. But if surrogate models are to replace DFT calculations in dynamical simulations, then a minimal NN architecture resulting in minimal inference times would be desirable. To this end, Fig.~\ref{fig:network_sizes} shows how the NNs identified by the different hyperparameter optimization methods differ greatly. The direct search favors a large NN, while the NASWOT mean score favors a small, single layer NNs. Optuna provides a middle ground, with a shallow NN, as does OAT. We deduce from Fig.~\ref{fig:hyperparameter_comparison_timings} that the Optuna NN is close to the globally optimal NN architecture. The NASWOT mean score captures this optimum to a sufficient degree, while providing a massive reduction in computation time which is further illustrated in Supplementary Note 1. Overall, all of the NNs identified by the considered methods are smaller than those predicted by the direct search algorithm, resulting in drastically decreased inference times. As can be seen in Fig.~\ref{fig:hyperparameter_comparison_timings}, these smaller networks also yield higher accuracy. One explanation for this is a more favorable optimization landscape, or problems with over-fitting for large models. 

\subsection{Descriptor surrogate metric}
While the aforementioned hyperparameter optimization is important to any conceivable ML problem that aims to map a vector quantity to another vector, there are, however, hyperparameters that are inherent to identifying suitable descriptors.
In our grid-based approach to learning the electronic structure, we rely on descriptors that encode the local atomic environment around a point in the simulation cell.
Since there is no clear physical relation between the NN prediction accuracy and the hyperparameters characterizing the way such a local environment is captured, data preprocessing itself requires a hyperparameter optimization. This requires the repeated training of a NN using a wide range of descriptors. The NASWOT mean score cannot be employed here, as it is not the network architecture we seek to optimize. 
We therefore introduce a \emph{descriptor surrogate metric} called ACSD which is based on similarity measures. It facilitates identifying the optimal choice of descriptors for particle-mesh data. Similar to NASWOT, it is highly efficient and achieves a speed-up of two orders of magnitude compared to conventional NN training, because it enables a training-free optimization.

In our workflow, we employ the SNAP descriptors~\cite{thompson_spectral_2015,wood_extending_2018,wood_data-driven_2019,cusentino_explicit_2020} to capture local atomic environments. However, local descriptors may be based on other established fingerprinting schemes for atomic configurations, such as SOAP~\cite{bartok_representing_2013}, the Coulomb matrix~\cite{rupp_fast_2012}, BoB~\cite{hansen_machine_2015}, FCHL~\cite{faber_alchemical_2017}, or ACE~\cite{drautz_atomic_2019, lysogorskiy_performant_2021}. Assuming that we investigate cells consisting only of one chemical species, SNAP descriptors are calculated via the local atomic density around a grid-point
\begin{equation}
    \rho(\bfr) = \delta(\mathbf{0})+\sum_{r_k<R_\mathrm{cut}}f_c(r_k;R_\mathrm{cut})\delta(\bm{r}_k) \; , \label{eq:SNAPDescriptors1}
\end{equation}
after placing said grid-point at the origin. In Eq.~(\ref{eq:SNAPDescriptors1}), $\bm{r}_k$ with $r_k=|\bm{r}_k|$ is the position of atom $k$ and $R_\mathrm{cut}$ is the cutoff radius which determines the length scale of the atomic environment considered by the SNAP descriptor. This local atomic density is represented in terms of four-dimensional hyperspherical coordinates. The number of terms in this expansion is denoted by $j$ with a dimensionality governed by $J_\mathrm{max}$; the higher $J_\mathrm{max}$, the more components per grid-point are taken into account.

Our proposed surrogate metric is based on analyzing a similarity measure between the output and input vectors of the NN, namely the SNAP vectors which are the input and the LDOS vectors which are the output. Given two points on the real space grid of a simulation cell $\bfr_1$ and $\bfr_2$, we compute the cosine similarity $S_C$ for either two LDOS vectors $d(\epsilon, \bfr)$ and two SNAP vectors $B(j, \bfr)$ as
\begin{equation}
    S_C(X_1, X_2) = \frac{X_1\cdot X_2}{\|{X_1}\| \|{X_2}\|} \label{eq:cosine_similarity}
\end{equation}
with $X_1=d(\epsilon,\bfr_1), X_2=d(\epsilon,\bfr_2)$ or $X_1=B(j,\bfr_1), X_2=B(j,\bfr_2)$. 
We then compute a 2D point cloud $\left\{S^{i}_C(B), S^{i}_C(d)\right\}$ with $S^{i}_C(B)=S_C\left\{B(j, \bfr_t), B(j, \bfr_s)\right\}$, $S^{i}_C(d)=S_C\left\{d(\epsilon, \bfr_t), d(\epsilon, \bfr_s)\right\}$ for $N_\mathrm{sim}$ points $\left\{t,s\right\}$ sampled from the simulation cell. We consider $N_\mathrm{sim}=200\times200=40,000$ points, i.e., for each grid point in a set of 200 grid points, these distances are determined w.r.t.~200 randomly drawn points.

The optimal choice of descriptors is determined by the hyperparameters $R_\mathrm{cut}$ and $J_\mathrm{max}$. We need to consider two limiting cases: (i) when $S^{i}_C(B) \ll S^{i}_C(d)$, even drastically dissimilar descriptors might yield the same LDOS, and modeling becomes trivial from a ML perspective. This case applies when $R_\mathrm{cut}$ is small.
However, this in turn means physically less informed descriptors and therefore lower prediction accuracy, because the length scale of the atomic environment is small. 
If we instead choose $R_\mathrm{cut}$ to be large, the descriptors will be physically well informed. But we risk approaching (ii) $S^{i}_C(B) \gg S^{i}_C(d)$, which makes our problem difficult to model in terms of ML. 
For a fixed choice of $R_\mathrm{cut}$, the number of expansion coefficients governed by $J_\mathrm{max}$ determines which of these limiting cases is approached. As we aim for optimal performance w.r.t.~both accuracy and data footprint, we argue that finding a combination of maximum $R_\mathrm{cut}$ and minimum $J_\mathrm{max}$ for which $S^{i}_C(B) \approx S^{i}_C(d)$ is the optimal choice. 

To judge whether such a combination has been found, we introduce the ACSD. It is defined as the average difference
\begin{equation}
    \mathrm{ACSD} = \frac{1}{N_\mathrm{sim}}\sum_i^{N_\mathrm{sim}} |S^i_C(d)-S^i_C(B)| \label{eq:ACSD}
\end{equation}
between all points within the 2D point cloud $\left\{S^{i}_C(B), S^{i}_C(d)\right\}$ and $\left\{S^{i}_C(B), S^{i}_C(B)\right\}$, where $N_\mathrm{sim}$ denotes the number of sample points. The ACSD thus measures the average over the distribution of similarities that deviate from the $S^{i}_C(B) = S^{i}_C(d)$ line. Eq.~(\ref{eq:ACSD}) can be evaluated rapidly in contrast to lengthy NN training required for a traditional hyperparameter search. To investigate the accuracy of the ACSD, we consider 20 snapshots of 128 beryllium atoms at room temperature (298 K) and ambient mass density (1.896 g/cc) \cite{fiedler_ldossnap_2022}.

\begin{figure}[ht]
    \centering
    \includegraphics[width=0.95\columnwidth]{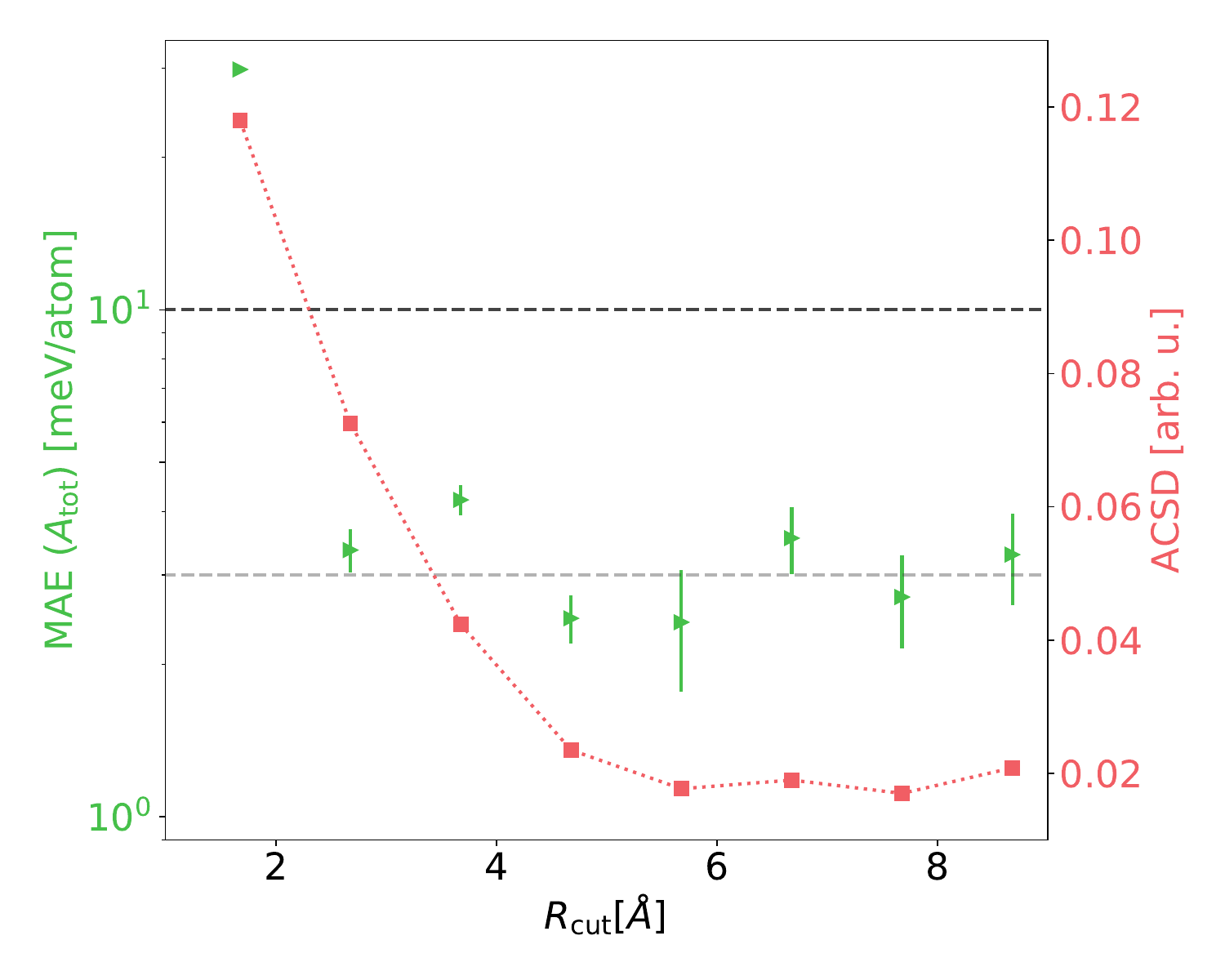}
    \caption{NN errors when using SNAP descriptors with differing $R_\mathrm{cut}$. $J_\mathrm{max}$ was kept at 5. The dark dashed line indicates an error of 10 meV/atom, the upper threshold a model should achieve to be comparable to other competetive models, while the lighter dashed line indicates an error of 3 meV/atom, which is the accuracy reported in Ref.~\cite{ellis_accelerating_2021}.}
    \label{fig:snap_analysis_cutoff_radius}
\end{figure}

\begin{figure}[ht]
    \centering
    \includegraphics[width=0.95\columnwidth]{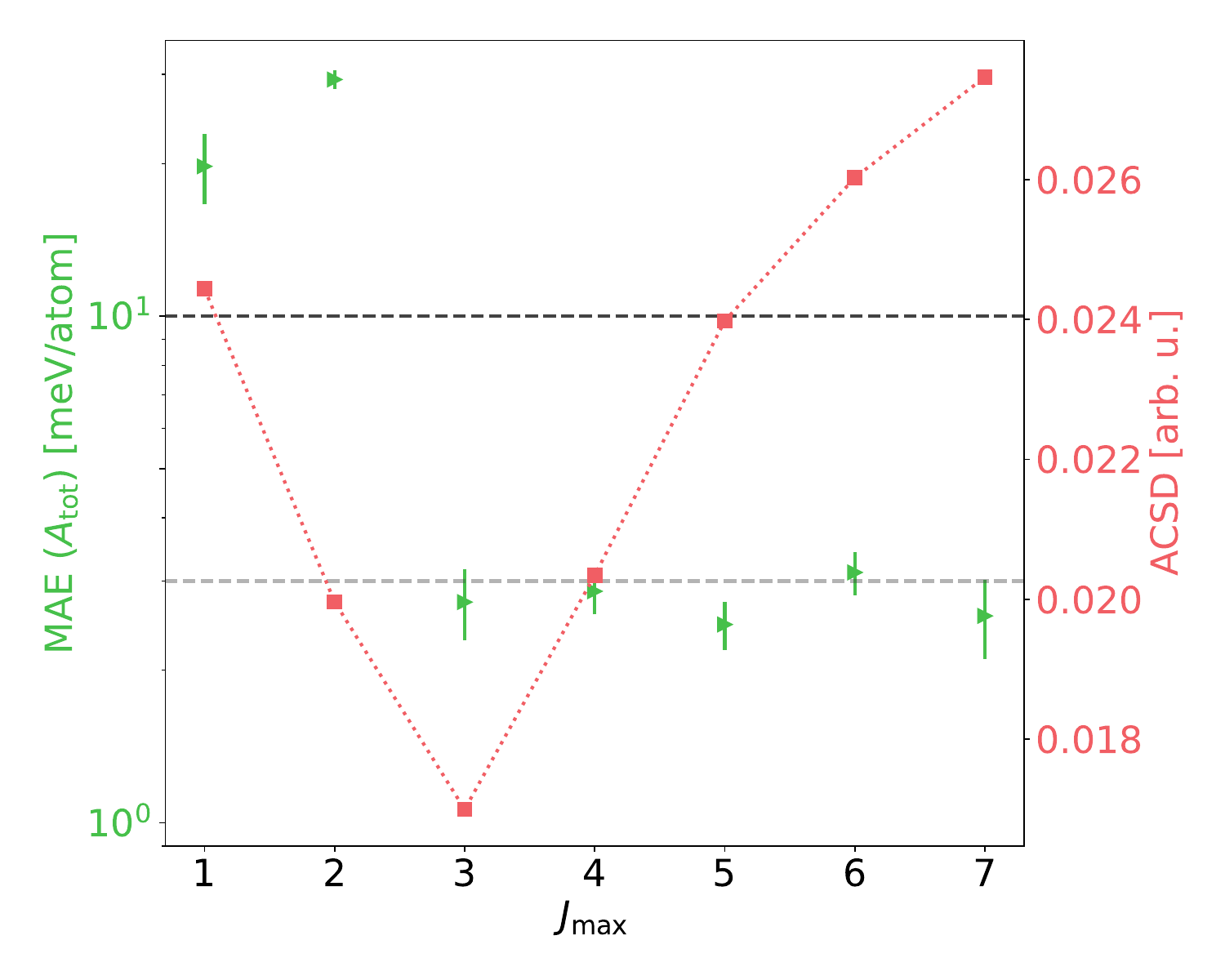}
    \caption{NN errors when using SNAP descriptors with differing $J_\mathrm{max}$. $R_\mathrm{cut}$ was kept at $4.676$\AA~. The dark dashed line indicates an error of 10 meV/atom, the upper threshold a model should achieve to be comparable to other competetive models, while the lighter dashed line indicates an error of 3 meV/atom, which is the accuracy reported in Ref.~\cite{ellis_accelerating_2021}.}
    \label{fig:snap_anaylysis_components}
\end{figure}

To assess the utility of the ACSD as a rapid and reliable \emph{descriptor surrogate metric}, we compare the predicted ACSD with actual MAE of the total energy inferred from the NN.
To that end we consider the hyperparameters $R_\mathrm{cut}=4.676$\AA~and $J_\mathrm{max}=5$ which were identified as accurate in Ref.~\cite{ellis_accelerating_2021}. We then vary both hyperparameters, one at a time with the other held fixed. The reference result was calculated for all these hyperparameter combinations. This involved generating SNAP descriptors, training the NNs, and predicting the total free energy based on these NNs. The central results of this assessment are shown in Fig.~\ref{fig:snap_analysis_cutoff_radius} for a fixed number of components $J_\mathrm{max}$ and varying cutoff radius $R_\mathrm{cut}$, and vice versa in Fig.~\ref{fig:snap_anaylysis_components}.

It is evident in Fig.~\ref{fig:snap_analysis_cutoff_radius} that the NN prediction (green) becomes more accurate with increasing $R_\mathrm{cut}$, up until a minimum at 4.676~\AA\,(achieving the desired accuracy of below 3 meV/atom), after which a slight decrease in both average accuracy and spread is observed. This behavior is reproduced, almost exactly, by the ACSD surrogate metric (red). 
These results follow the intuitive expectation: a small $R_\mathrm{cut}$ means that individual SNAP descriptors carry less information about the atomic environment, making it harder for a NN to actually predict the electronic structure from the data provided. On the other hand, very large values of $R_\mathrm{cut}$ lead to SNAP descriptors that incorporate information from almost the entire simulation cell. These tend to be similar to one another, even though the actual electronic structure at a particular grid point differs, complicating training. Consequently, we expect the optimal $R_\mathrm{cut}$ in between these extremes, which is correctly identified by the ACSD. Furthermore, Fig.~\ref{fig:snap_analysis_cutoff_radius} confirms the prior assertion that ML modeling becomes increasingly more challenging as $R_\mathrm{cut}$ increases, as is evident from the increasing spread in the network prediction errors. 

The computational speed-up of this method is drastic. The conventional method of finding optimal descriptors is a computationally heavy task, because it requires multiple NN model optimizations. Contrarily, the evaluation of the ACSD surrogate metric can be done in a matter of minutes on a single CPU. This results in a speed-up of around two orders of magnitude, while qualitatively yielding the same results.

The assessment for a fixed cutoff radius $R_\mathrm{cut}$ and varying $J_\mathrm{max}$ yields a similar trend which is illustrated in Fig.~\ref{fig:snap_anaylysis_components}. As long as $J_\mathrm{max}$ is chosen sufficiently large, there is little effect on the accuracy of the NN (green), and above $J_\mathrm{max}=2$ all networks achieve the desired accuracy. This trend is not fully reflected by the ACSD surrogate metric (red). However, this can be explained due to the nature of this numerical experiment. An increase in components leads to additional components being added that are generally small in value. These in turn cause slightly larger deviations for almost identical SNAP vectors, but do not carry meaningful information for the NN. This leads to almost unnoticeable differences in NN accuracies. However, minimal $J_\mathrm{max}$ values leading to reasonable accuracy is indeed reflected by the ACSD surrogate metric. Based only on the surrogate metric, one would choose the first data point for which the ACSD encounters a minimum, in this case $J_\mathrm{max}=3$. The NN accuracy of this data set is comparable to that for higher $J_\mathrm{max}$ at reduced computational cost, thus demonstrating the utility of our ACSD surrogate metric.

\section{Discussion and Outlook}
We tackle the complex optimization problem inherent to any NN surrogate model construction in the context of electronic structures. This constitutes a major computational bottleneck in addition to data generation. 
We provide a highly efficient and automated hyperparameter optimization workflow for generating ML surrogate models of electronic structures. It speeds up this process by two orders of magnitude (see Fig.~\ref{fig:full_workflow}), as we demonstrate in terms of a comprehensive comparison with state-of-the-art techniques. Our workflow consists of two advances -- a \emph{training-free score} for rapid hyperparameter optimization of NNs and a \emph{descriptor surrogate metric} enabling an efficient search for suitable descriptors of particle-mesh data. 

We have first assessed the accuracy and efficiency of our workflow against multiple hyperparameter optimization techniques such as the direct search, OAT, and the standard Optuna library. 
We achieve large gains in computational efficiency in hyperparameter optimization by adapting the NASWOT method \cite{mellor_neural_2020} as a NASWOT mean score into our workflow. This method does not require any NN training up until an optimal set of hyperparameters has been identified. With our NASWOT mean score we are able to calculate a surrogate model in a few hours, whereas the state-of-the-art methods take days.
In addition, if higher accuracy is needed, we showed that combining Optuna with the NASWOT mean score outperforms traditional search approaches. 
We also found that our hyperparameter optimization workflow impacts model performance. NN architectures identified as optimal by the NASWOT mean score algorithm are equally robust as the direct search, but yield smaller NNs with optimal inference performance. 

Furthermore, we have developed the ACSD \emph{descriptor surrogate metric} to find hyperparameters for the calculation of suitable particle-mesh descriptors without having to train any NN models. Likewise, our algorithm speeds up the state of the art by two orders of magnitude. First research into applying our models at larger temperatures suggest that the ACSD successfully recovers the expected physicys, i.e.~favoring smaller cutoff radii (reflecting the larger disorder in the system). 

By incorporating these two developments, we have devised a highly efficient ML surrogate modelling workflow shown in Fig.~\ref{fig:full_workflow}. All steps in this workflow can easily be automated with the algorithms considered here.
These tools will enable a breadth of future applications in which large parts of the data processing for DFT surrogate models can be automated or executed with minimal user input. Our final workflow reduces the time required to construct surrogate models by two orders of magnitude. It thus provides a pathway to employing DFT surrogate models in large-scale investigations of materials under a variety of conditions.

\section*{Acknowledgments}

A.C. acknowledges useful discussions with Michael Bussmann. 
L.F. thanks Alexander Debus for providing us with additional computing time. 
We gratefully acknowledge computation time on the Bull Cluster at the Center for Information Services and High Performace Computing (ZIH) at Technische Universität Dresden.
Sandia National Laboratories is a multimission laboratory managed and operated by National Technology \& Engineering Solutions of Sandia, LLC, a wholly owned subsidiary of Honeywell International Inc., for the U.S. Department of Energy’s National Nuclear Security Administration under contract DE-NA0003525. This paper describes objective technical results and analysis. Any subjective views or opinions that might be expressed in the paper do not necessarily represent the views of the U.S. Department of Energy or the United States Government.
This work was partially supported by the Center of Advanced Systems Understanding (CASUS) which is financed by Germany’s Federal Ministry of Education and Research (BMBF) and by the Saxon state government out of the State budget approved by the Saxon State Parliament.

\section*{References}
\bibliographystyle{iopart-num}
\providecommand{\newblock}{}

\end{document}